\documentclass{PoS}

\newcommand{\kslash}{k\kern-1ex /}
\newcommand{\pslash}{p\kern-1ex /}
\newcommand{\qslash}{q\kern-1ex /}
\newcommand{\lslash}{l\kern-1ex /}
\newcommand{\sslash}{s\kern-1ex /}
\newcommand{\Dslash}{{\cal D}\kern-1.5ex /}

\newcommand{\beqa}{\begin{eqnarray}}
\newcommand{\eeqa}{\end{eqnarray}}

\newcommand{\be}{\begin{equation}}
\newcommand{\ee}{\end{equation}}
\newcommand{\ben}{\begin{eqnarray}}
\newcommand{\een}{\end{eqnarray}}
\newcommand{\bd}{\begin{description}}
\newcommand{\ed}{\end{description}}

\def\lsim{\raise0.3ex\hbox{$<$\kern-0.75em\raise-1.1ex\hbox{$\sim$}}}
\def\gsim{\raise0.3ex\hbox{$>$\kern-0.75em\raise-1.1ex\hbox{$\sim$}}}
\def\simgt{\rlap{\lower 3.5 pt\hbox{$\mathchar \sim$}}\raise 1pt \hbox {$>$}}
\def\simlt{\rlap{\lower 3.5 pt\hbox{$\mathchar \sim$}}\raise 1pt \hbox {$<$}}
\def\Simgt{\rlap{\lower 6.5 pt\hbox{$\mathchar \sim$}}\raise 3pt \hbox {$>$}}
\def\Simlt{\rlap{\lower 6.5 pt\hbox{$\mathchar \sim$}}\raise 3pt \hbox {$<$}}

\newcommand{\la}{\langle}
\newcommand{\ra}{\rangle}

\rightline{\sffamily UTCCS-P-22, HUPD-0606}
\vspace*{-5mm}

\title{2+1 Flavor Lattice QCD with L{\" u}scher's Domain-Decomposed
HMC Algorithm}
\ShortTitle{2+1 Flavor Lattice QCD with LDDHMC Algorithm}

\author{PACS-CS Collaboration:
 \speaker{Y.~Kuramashi}${}^{a,b}$\thanks{E-mail: kuramasi@het.ph.tsukuba.ac.jp}~,
 S.~Aoki${}^{a,c}$,
 K.-I.~Ishikawa${}^{d}$,
 T.~Ishikawa${}^{b}$,
 N.~Ishizuka${}^{a,b}$,
 K.~Kanaya${}^{a}$,
 N.~Tsutsui${}^{e}$,
 M.~Okawa${}^{d}$,
 Y.~Taniguchi${}^{a,b}$,
 A.~Ukawa${}^{a,b}$,
 T.~Yoshi\'e${}^{a,b}$,
 \\
 \llap{${}^a$}Graduate School of Pure and Applied Sciences, University of Tsukuba, Tsukuba, Ibaraki 305-8571, Japan\\
 \llap{${}^b$}Center for Computational Sciences, University of Tsukuba, Tsukuba, Ibaraki 305-8577, Japan\\
 \llap{${}^c$}Riken BNL Research Center, Brookhaven National Laboratory, Upton, New York 11973, USA\\
 \llap{${}^d$}Department of Physics, Hiroshima University, Higashi-Hiroshima, Hiroshima 739-8526, Japan\\
 \llap{${}^e$}High Energy Accelerator Research Organization (KEK),
               Tsukuba 305-0801, Japan}
\abstract{We report on a study of 2+1 flavor lattice QCD
with the $O(a)$-improved Wilson quarks on a $16^3\times 32$
lattice at the lattice spacing $1/a\approx 2$GeV
employing L{\" u}scher's domain-decomposed HMC(LDDHMC) algorithm.
This is dedicated to a preliminary study for the PACS-CS project which
plans to complete the Wilson-clover $N_f=2+1$ program
lowering the up-down quark masses close to the physical values
as  much as possible.
We focus on three issues: (i) how light quark masses we can
reach with LDDHMC, (ii) efficiency of the algorithm compared with
the conventional HMC, (iii) parameter choice for the 
production runs on PACS-CS.
}

\FullConference{XXIVth International Symposium on Lattice Field Theory\\
		 July 23-28, 2006\\
		 Tucson, Arizona, USA}

\begin{document}

%1
%=================
\section{Introduction}
%=================
%
In the past decade the progress of simulation algorithms and 
the availability of more affluent computational resources have enabled
us to investigate two-  and three-flavor dynamical quark effects 
with lighter quark masses.
The CP-PACS and JLQCD joint project\cite{tishikawa} has performed 
a 2+1 flavor full QCD simulation employing the $O(a)$-improved Wilson 
quark action with nonperturbative $c_{\rm SW}$ 
and the Iwasaki gauge action\cite{iwasaki}
on a (2fm)$^3$ box at three lattice spacings.  
Unfortunately, the lightest up and down quark mass reached is
about 64MeV corresponding to $m_\pi/m_\rho\approx 0.6$. 

A goal of the next decade should be 
the realistic full QCD simulation with
the physical up, down and strange quark masses.
The PACS-CS project, which is based 
on the PACS-CS (Parallel Array Computer
System for Computational Sciences) computer
installed at University of Tsukuba on July 1 of 2006\cite{ukawa}, 
aims at this goal succeeding the Wilson-clover 2+1 program of
the previous CP-PACS/JLQCD project.
%We wish to reach the physical up and down quark masses. 

In this report we present a preparatory study for the PACS-CS project
using (1.6fm)$^3$ lattice. To simulate the dynamical up and down quarks,
we employ the domain-decomposed HMC algorithm proposed 
by L{\"u}scher\cite{lddhmc}. The effectiveness of this algorithm  
for small quark mass region is already shown in
the two-flavor case\cite{lddhmc,giusti}. The strange quark is included by the 
exact Polynomial HMC (PHMC) algorithm\cite{phmc}.
Our primary purpose is to investigate how light up and down quark masses
we can go down to
with LDDHMC. The efficiency of LDDHMC is compared to that of the
conventional HMC. Based on this study we finally 
make a parameter choice for the
production runs on PACS-CS, and briefly discuss the physics plan.

%2
%=================================
\section{Simulation details}
%=================================
%\subsection{Simulation parameters}

We employ the $O(a)$-improved Wilson  
quark action with nonperturbative $c_{\rm SW}$\cite{csw_np}
and the Iwasaki gauge action at $\beta=1.9$
on a $16^3\times 32$ lattice.
The lattice spacing is about 0.1fm from the previous CP-PACS/JLQCD
results on a $20^3\times 40$ lattice\cite{tishikawa}.
In Table~\ref{tab:param} we summarize 
the details of the simulation parameters.
We employ four degenerate up and down quark masses based on 
the previous CP-PACS/JLQCD results, while 
the strange quark mass is fixed at $\kappa_{\rm s}=0.1364$  
close to the physical point $\kappa_{\rm s}=0.136412(50)$\cite{tishikawa}. 
The LDDHMC algorithm is implemented with a $8^4$ block size, while
the exact PHMC algorithm is not domain-decomposed.
We choose the trajectory length of $\tau=0.5/\sqrt{2}$. 
With the integers $N_0$, $N_1$, $N_2$ the step sizes are given
by $\delta\tau=\tau/(N_0N_1N_2)$ for the gauge part,
$\delta\tau=\tau/(N_1N_2)$ for the strange quark and the
UV part of the up and down quarks, and $\delta\tau=\tau/N_2$
for the IR part of the up and down quarks.
The Wilson quark matrix inversion is carried out by the even/odd
preconditioned BiCGStab solver with the stopping condition 
$|Dx-b|/|b|<10^{-9}$ for the force calculation and 
$10^{-14}$ for the Hamiltonian.
%Before thermalization we discard 1500-2000 trajectories 
%depending on the up and down quarks.

All the simulations presented in this report are carried out
on Hitachi SR11000/J1 at Information Technology Center 
of the University of Tokyo.

\begin{table}
\centering{
\caption{Simulation parameters and basic physical quantities. 
\#mult is the number of multiplications of the Wilson-Dirac quark matrix
on the full lattice including even and odd sites.} 
%$P_{\rm acc}$(GMP) is analyzed with bin-size=1 traj.}
\label{tab:param}
\newcommand{\cc}[1]{\multicolumn{1}{c}{#1}}
\begin{tabular}{ccccc}
\hline
$\kappa_{\rm ud}$ & 0.13700 & 0.13741 & 0.13759  & 0.13770   \\
\hline
$N_0,N_1,N_2$   & 4,5,6 & 4,5,8 & 4,5,12 & 4,5,14 \\
$N_{\rm poly}$  & 130   &  140  & 140    & 140 \\
%CPU time[hrs]   & 23.0 & 35.2 & 60.3 & 79.6  \\
thermalization  & 2000 & 1700 & 1600 & 1500  \\
No. traj.       & 2000 & 2000 & 2000 & 900  \\
bin size[trajs] &   50   &  50  &  50  &  50 \\
$\la P\ra$   & 0.569126(35) & 0.569921(37) & 0.570341(43) & 0.570574(79)\\
%$\la dH\ra$  & 0.048(6) & 0.091(21) & 4.9(3.6) & 4.2(2.8) \\
%$R(|dH|>1)$   & 0.0015 & 0.027  & 0.0385 & 0.032 \\
$P_{\rm spike}(|dH|>2)$   & 0      & 0.002  & 0.018  & 0.017 \\
%$R(|dH|>5)$   & 0      & 0.0005 & 0.0125 & 0.013 \\
$P_{\rm spike}(|dH|>10)$  & 0      & 0.0005 & 0.0105 & 0.008 \\
$P_{\rm spike}(|dH|>100)$ & 0      & 0      & 0.002  & 0.003 \\
$\la {\rm e}^{-dH}\ra$  & 1.003(6) & 0.9995(99) & 0.9867(92) & 0.975(13) \\
$P_{\rm acc}$(HMC) & 0.88(1) & 0.86(1) & 0.89(1) & 0.86(2) \\
$P_{\rm acc}$(GMP) & 0.94(1) & 0.95(1) & 0.94(1) & 0.93(1) \\
\#mult/traj & 20284(61) & 31776(171) & 53403(395) & 69342(990) \\
%\#mult/traj(IR) & 6297 & 15181 & 28028 & 48274 \\
%\#mult/traj(poly) & 3900 & 5600 & 8400 & 9800 \\
$\tau_{\rm int}[P]$ & 10.6(2.7) & 16.0(6.6) & 18.0(6.0) & 36.8(23.6)  \\
$\tau_{\rm int}[P]\cdot \#{\rm mult}/10^3$ 
& 215(55) & 508(210) & 961(320) & 2552(1636) \\
%$|F^{(0)}|$ & 7.21550(14) & 7.21728(14)  & 7.21865(13) & 7.21907(26)\\
%$|F^{(1)}|$ & 1.95634(13) & 1.96989(15)  & 1.97444(20) & 1.97968(36) \\
%$|F^{(2)}|$ & 0.21100(23) & 0.2403(10)   & 0.2656(29)  & 0.2877(36) \\
\hline
No. config.         & 200 & 200 & 200 & 90  \\
bin size[configs]   &  5  &  5  &  5  &  5  \\
%bin size[trajs/10]  &  5  &  5  &  5  &  5  \\
fitting range$[t_{\rm min}, t_{\rm max}]$ 
& [8, 13] & [8, 13] & [7, 12] & [7, 12] \\
$m_{PS}(\kappa_{\rm ud})$ 
& 0.3303(21) & 0.2509(29)  & 0.185(5)   & 0.158(8) \\
%$m_V(LL)$        & 0.529(5)   & 0.490(12)   & 0.431(13)  & 0.384(21) \\
%$m_{PS}/m_V(LL)$ & 0.624(7)   & 0.512(11)   & 0.429(16)  & 0.413(22) \\
%$m_{PS}(SS)$     & 0.4073(16) & 0.3959(20)  & 0.3731(24) & 0.3666(31) \\
%$m_V(SS)$        & 0.577(4)   & 0.566(4)    & 0.537(3)   & 0.530(7) \\
%$m_{PS}/m_V(SS)$ & 0.706(5)   & 0.699(5)    & 0.695(4)   & 0.692(6) \\
$m_{PS}(\kappa_{\rm ud})$[MeV] 
& 655(4) & 498(6)  & 367(9)   & 313(16) \\
$m_{ud}^{\rm AWI}$[MeV] & 63.7(4) & 34.8(5) & 20.7(5)    & 15.2(12) \\
$m_{\rm PS}^2/m_{\rm ud}^{\rm AWI}$ & 3.39(3) & 3.59(7) 
& 3.27(14) & 3.27(23) \\
\hline
No. config.         & 200 & 200 & 200 & 90  \\
bin size[configs]   &  5  &  5  &  5  &  5  \\
%bin size[trajs/10]  &  5  &  5  &  5  &  5  \\
$\mu_{\rm median}$  & 0.02018     & 0.01077     & 0.00718     & 0.00602 \\
$\mu_{\rm average}$ & 0.01982(24) & 0.01059(22) & 0.00716(14) & 0.00619(26) \\
$\sigma$            & 0.00209(16) & 0.00203(15) & 0.00166(15) & 0.00138(21) \\
$\sigma\sqrt{V}/a$  & 0.755(60)   & 0.735(53)   & 0.601(54)   & 0.501(76) \\
\hline
\end{tabular}
%\\[2pt]
}
\end{table}

\section{Numerical results}

\subsection{History of $dH$}

\begin{figure}[t]
\vspace{3mm}
\begin{center}
\includegraphics[width=110mm,angle=0]{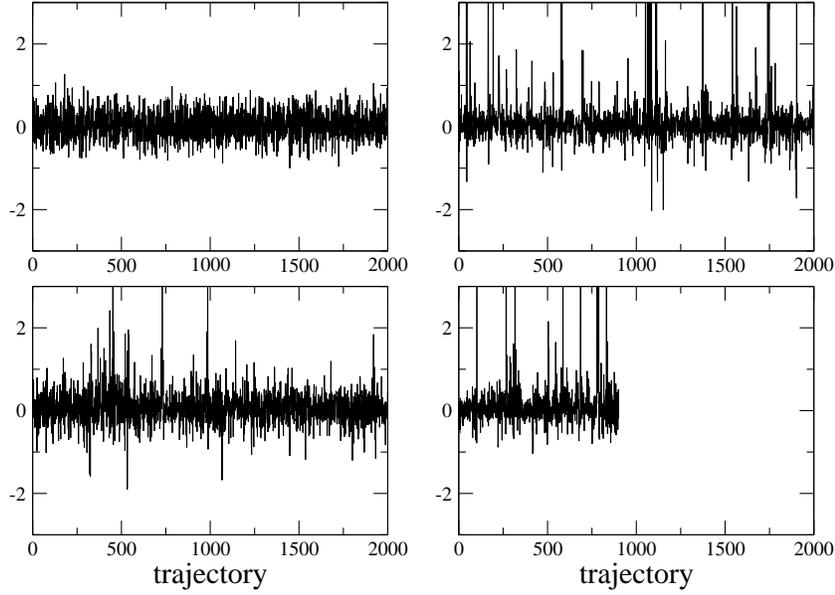}
\end{center}
\vspace{-.5cm}
\caption{Time histories of $dH$ as a function of the trajectory
number after thermalization for $\kappa_{\rm ud}=0.13700$(top left),
$0.13741$(bottom left), 
$0.13759$(top right) 
and $0.13770$(bottom right).}
\label{fig:dh}
\end{figure}

In Fig.\ref{fig:dh} we show the history of $dH$  for 
each $\kappa_{\rm ud}$. We observe a frequent occurrence of spikes for
$\kappa_{\rm ud}=0.13759$ and 0.13770, whose
probabilities for $|dH|>2$ are about 2\%.
Although the acceptance ratios  for
$\kappa_{\rm ud}=0.13759$ and 0.13770
are still sufficiently high, 
$\la {\rm e}^{-dH}\ra$ which measures the area preserving property
deviates from unity by 2$\sigma$.
It is clear that the spikes are a potential source to
violate the area preserving property.
In order to avoid any problem associated with the spikes,
we shall incorporate the replay trick\cite{lddhmc, replay}
in the production run.

\subsection{Magnitude of force terms}

\begin{figure}[t]
\begin{center}
\includegraphics[width=70mm,angle=0]{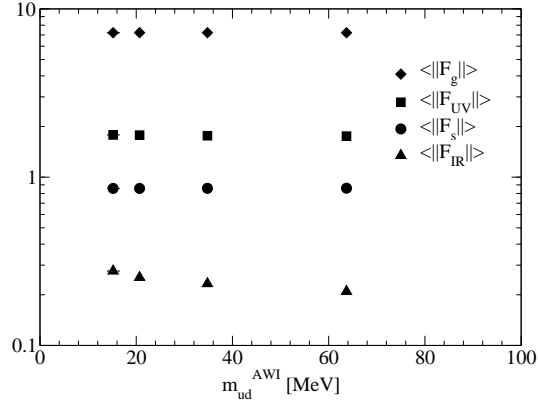}
\end{center}
\vspace{-.5cm}
\caption{Magnitude of each force term as a function of the AWI 
quark mass $m_{\rm ud}^{\rm AWI}$. The errors are within symbols.}
\label{fig:force}
\end{figure}

Figure~\ref{fig:force} shows the magnitude of the four force terms as a
function of the axial vector Ward identity quark mass $m_{\rm ud}^{\rm AWI}$ 
discussed below; $F_g$ denotes the gauge part, $F_{\rm UV, IR}$ for
the UV and IR parts of the up and down quarks, and $F_s$ for 
the strange quark.   
Note that the definition of the magnitude of force term is a factor two
larger than that in Ref.\cite{lddhmc}.
We observe that $\|F_g\|$, $\|F_{\rm UV}\|$ and  $\|F_s\|$
are almost independent of the quark mass, while $\|F_{\rm IR}\|$ gradually
increases as the up and down quark masses decreases.
At $\kappa_{\rm ud}=0.13770$ we find
\ben
\|F_g \|:\|F_{\rm UV} \|:\|F_{\rm s} \|:\|F_{\rm IR} \|\approx
24:6:3:1.
\een  
This result suggests that
our choice of $\delta\tau$ for the strange quark may be unnecessarily fine.

\subsection{Comparison of LDDHMC and HMC}

For a direct comparison between LDDHMC and HMC 
we have repeated the 2+1 flavor simulation at $\kappa_{\rm
ud}=0.13700$ and $\kappa_{\rm
s}=0.13640$  with the conventional
HMC algorithm employing the same parameters
except $\tau=1$ for HMC. 
Analyzing 3000 trajectories after thermalization we obtain 
$\tau_{\rm int}[P]=6.6(1.8)$ and \#mult/traj=148625(626) for HMC.
Since the trajectory lengths are differently chosen for LDDHMC and HMC,
we compare their efficiency by 
$\tau_{\rm int}[P]\cdot \#{\rm mult}/10^3$ which should be 
independent of the trajectory length.
This quantity is found to be 
215(55) for LDDHMC and 981(268) for HMC.
Hence LDDHMC is almost five times 
more efficient than 
HMC at $m_{\rm PS}/m_{\rm V}\approx 0.6$.
%However, 
We should keep in mind that this comparison involves
the strange quark part simulated by the exact PHMC algorithm.

\subsection{Quark masses}

The up and down quark masses are
measured by using the axial vector Ward identity (AWI):  
\ben
m_{\rm ud}^{\rm AWI}=\frac{Z_A}{Z_P}
\frac{\langle 0|\nabla_4 A_4^{\rm impr}|PS\rangle}
{2 \langle 0|P|PS\rangle},
\een
where the renormalization factors $Z_{A,P}$ and the improvement
coefficients are determined perturbatively up to one-loop level.
The results are given in Table~\ref{tab:param} in physical units
together with the corresponding pseudoscalar masses.
It is encouraging that LDDHMC allows a simulation 
at $m_{\rm ud}^{\rm AWI}=15$MeV which is roughly a quarter of 
the lightest quark mass employed in the previous CP-PACS/JLQCD
project. 
In Table~\ref{tab:param} we also give the results for
$m_{\rm PS}^2/m_{\rm ud}^{\rm AWI}$, whose small quark mass dependence 
indicates that the finite volume effects are not sizable.

\subsection{Spectral gap distribution} 

Recently Del~Debbio {\it et al.} have discussed the relevance of the spectral 
gap distribution of the Wilson-Dirac
operator for the stability of two-flavor lattice QCD simulations\cite{gap}. 
This study is applied to our three flavor case. 
We are interested in the effects of the
strange quark contributions.

\begin{figure}[t]
\begin{center}
\includegraphics[width=110mm,angle=0]{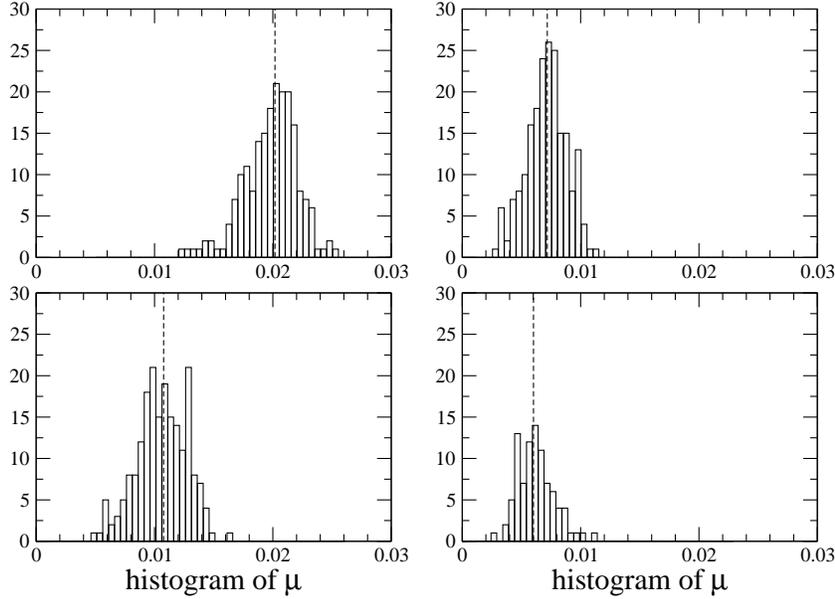}
\end{center}
\vspace{-.5cm}
\caption{Histograms of $\mu$ for $\kappa_{\rm ud}=0.13700$(left top),
$0.13741$(left bottom), 
$0.13759$(right top) 
and $0.13770$(right bottom). 
Vertical dashed lines denote $\mu_{\rm median}$.}
\label{fig:gap}
\end{figure}

\begin{figure}[t]
\vspace{3mm}
\begin{center}
\includegraphics[width=70mm,angle=0]{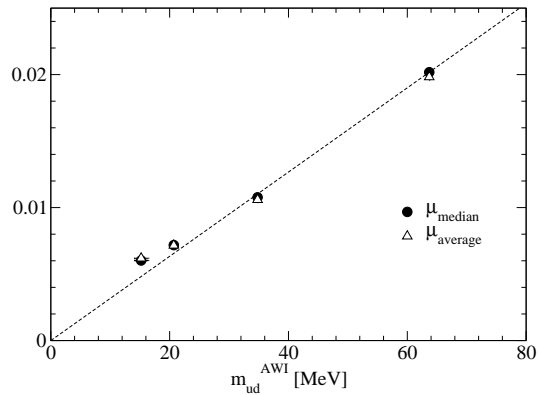}
\end{center}
\vspace{-.5cm}
\caption{Median and average of the gap distribution as a function of 
AWI quark mass $m_{\rm ud}^{\rm AWI}$. The dashed line drawn from the
origin to $\mu_{\rm median}$ at the largest quark mass is to guide the eye.}
\label{fig:mu}
\end{figure}

Following Ref.\cite{gap} we use the hermitian operator 
$Q_m=\gamma_5 D_m$ rather than the Wilson-Dirac operator 
$D_m=(1/2)\{\gamma_\mu(\nabla_\mu^*+\nabla_\mu)-a\nabla_\mu^*\nabla_\mu\}+m_0$.
The spectral gap is defined as
\ben
\mu={\rm min}\{|\lambda|\; |\; 
\lambda\;\mbox{is an eigenvalue of}\;Q_m\}.
\een
In Fig.\ref{fig:gap} we show histograms of the spectral gap $\mu$
for our four values of $\kappa_{\rm ud}$, which are obtained by the
implicitly restarted Lanczos algorithm\cite{lanczos}. 
We observe roughly symmetric distributions for all the quark masses.
Their median and average given   
in Table~\ref{tab:param} are consistent with each other 
within error bars.
In Fig.\ref{fig:mu} we plot the quark mass dependencies for 
$\mu_{\rm median}$ and $\mu_{\rm average}$.
Although both quantities are roughly
proportional to the AWI quark mass, we find  
a clear tendency that they 
deviate upwards from the linearity in terms of $m_{\rm ud}^{\rm AWI}$ 
toward the chiral limit. 

For the width of the distribution $\sigma$ we employ the same definition as
in Ref.\cite{gap}: $\sigma$ is defined as 
$(v-u)/2$, where $[u,v]$ is the smallest range of $\mu$ which contains
more than 68.3\% of the data. This is to avoid potentially large
statistical uncertainties which might occur when data are not sufficiently 
sampled.
The results of $\sigma$ are given in Table~\ref{tab:param}.
We find that the width of the distribution diminishes 
as the up and down quark mass decreases.
An intriguing quantity is the combination
$\sigma\sqrt{V}/a$: In the two-flavor case Del~Debbio {\it et al.} found 
that its value is roughly unity independent of the quark mass,
volume, and lattice spacing\cite{gap}. 
In our three-flavor case, on the contrary, we find $\sigma\sqrt{V}/a<1$
and its value decreases as the up and down quark mass decreases.
This fact may suggest that the contributions of the strange quark stabilize
the simulation at the lighter quark masses by shrinking
the gap distribution of the Wilson-Dirac operator.

%========================================
\section{Plan for production run on PACS-CS}
%========================================

%\subsection{Simulation parameters}
%Based on the study at $\beta=1.90$ with $16^3\times 32$
%we make a parameter choice for the production run on PACS-CS.

We employ the LDDHMC algorithm with the replay trick 
for the up and down quarks. The Wilson-Dirac quark matrix
inversion on the full lattice is accelerated by
the SAP+GCR solver\cite{sap+gcr}, which is three times faster
than the BiCGStab algorithm for our parameter range\cite{sap+gcr}. 
For the strange quark we use the UV-filtered PHMC algorithm,
which is two to three times more efficient than the PHMC 
algorithm\cite{uvphmc}.

We choose $\beta=1.83$, 1.90, 2.05 which are the same as
in the previous CP-PACS/JLQCD project\cite{tishikawa}.
The physical lattice volume is fixed at (3.0fm)$^3$:
$24^3\times 48$ for $\beta=1.83$, 
$32^3\times 64$ for $\beta=1.90$ and 
$40^3\times 80$ for $\beta=2.05$.
Assuming that the relation $\sigma\propto a/\sqrt{V}$ found in Ref.\cite{gap}
for the two-flavor simulation is applicable to the three-flavor case,
we expect that a stable simulation at $m_{\rm ud}^{\rm AWI}=4$MeV is possible
at $\beta=1.90$ and even smaller up and down quark masses are accessible
at $\beta=2.05$. 
Two strange quark masses are employed to interpolate the results
on the physical point.
We plan to accumulate 100 independent configurations with $10^4$ trajectories.

%\subsection{Physics plan}

The first step of the project is a simulation 
with the up and down quark mass down to
$m_{\rm ud}^{\rm AWI}=7$MeV, which is about twice the physical mass and
one order smaller than the lightest quark mass employed in the previous 
CP-PACS/JLQCD project. Once this is achieved,
we can almost remove the systematic errors associated with the 
chiral extrapolation which are currently one of the most significant errors
in lattice QCD simulations.
The hadron spectrum together with the quark masses 
should be determined precisely. We are now preparing a nonperturbative 
renormalization factor for the quark masses 
employing the Schr{\"o}dinger functional method.

There are two physical subjects which are especially interesting
for the light quark masses.
One is physics associated with topology.
The CP-PACS/JLQCD Collaborations presented an encouraging result 
for the $\eta^\prime$ meson mass\cite{eta}. 
An intriguing point is that 
the signals become cleaner as the up and down quark mass decreases.
The other is the $I=0$ $\pi$-$\pi$ scattering.
In Ref.\cite{i=0} we discussed that the light dynamical quarks
are essential ingredients to study it.

%========================================
%\section{Conclusion and outlook}
%========================================

%=====================
%\section*{Acknowledgments}
%=====================
%

This work is supported in part by Grants-in-Aid for Scientific Research 
from the Ministry of Education, Culture, Sports, 
Science and Technology (Nos.~13135204,
13135216,
15540251,
16540228,
16740147,
17340066,
17540259,
%17740171,
18104005,
18540250,
18740130).
%18740139).

\end{document}